\documentclass[a4,11pt]{cip-v3-cls-submit}
\usepackage{mathptmx}
\DeclareSymbolFont{epsilon}{OML}{cmm}{m}{it}
\DeclareMathSymbol{\epsilon}{\mathord}{epsilon}{"0F}
\usepackage{hyperref}
\hypersetup{
	colorlinks=false,
	pdfborder={0 0 0}
}
\usepackage{graphics,graphicx,subfigure,wrapfig,epstopdf}
\usepackage{wasysym,amsmath,amsxtra,amssymb,latexsym,amscd,color,cite,bm,array,multirow}
\usepackage[mathscr]{eucal}
\def\authors#1{\author{\begin{flushleft}{#1}\end{flushleft}}}
\def\authord#1#2{\textbf{\indent{#1}$^{#2}$}}
\def\addressed#1#2{\\[1mm]\textit{$\!\!\!^{#1}$\indent#2}}
\def\CorrEmail#1{\\[4mm]
	\textit{E-mail:}~$^\dag${#1}}

\def\Keywords#1{$\qquad$\\[-.35cm] \textnormal{Keywords:~{#1}}.} 

\def\and{\textbf{and} }
\def\Classification#1{$\quad$\\[-.35cm] \textnormal{Classification numbers:~{#1}.}}

\newcommand{\eq}[1]{equation~(\ref{#1})}
\newcommand{\eqs}[2]{equations~(\ref{#1}--\ref{#2})} 

\begin{document}

\Year{2025}
	\title{Energetic spectra from semi-implicit particle-in-cell simulations of magnetic reconnection}
        \authors{
	\authord{K. M. Schoeffler}{1,\dagger}, \authord{F. Bacchini}{2,3}, \authord{K. Kormann}{1}, \authord{B. Eichmann}{1}
	\newline
\and \authord{M. E. Innocenti}{1}
	\newline
	\addressed{1}{Institut f\"ur Theoretische Physik, Ruhr-Universit\"at Bochum, Bochum, Germany}
	\addressed{2}{Centre for mathematical Plasma Astrophysics, Department of Mathematics, KU Leuven, Leuven, Belgium}
        \addressed{3}{Royal Belgian Institute for Space Aeronomy, Solar-Terrestrial Centre of Excellence, Uccle, Belgium}    
\CorrEmail{Kevin.Schoeffler@rub.de}
}
\maketitle
	\markboth{semi-implicit PIC simulations of magnetic reconnection}{Schoeffler \textit{et al.}}

\begin{abstract}
Astrophysical observations suggest that magnetic reconnection in relativistic plasmas plays an important role in the acceleration of energetic particles.  Modeling this accurately requires numerical schemes capable of addressing large scales and realistic magnetic field configurations without sacrificing the kinetic description needed to model particle acceleration self-consistently.  We demonstrate the computational advantage of the relativistic semi-implicit method (RelSIM), which allows for reduced resolution while avoiding the numerical instabilities typically affecting standard explicit methods, helping to bridge the gap between macroscopic and kinetic scales. Two- and three-dimensional semi-implicit particle-in-cell simulations explore the linear tearing instability and the nonlinear development of reconnection and subsequent particle acceleration starting from a relativistic Harris equilibrium with no guide field.  The simulations show that particle acceleration in the context of magnetic reconnection leads to energetic power-law spectra with cutoff energies, consistent with previous work done using explicit methods, but are obtained with a considerably reduced resolution.
\end{abstract}

\Keywords{plasma, reconnection, tearing, particle-in-cell, semi-implicit, simulation}

\Classification{52.35.Vd, 95.30.Qd, 52.27.Ny, 52.65.-y, 52.65.Rr}

\section{Introduction}\label{sec:intro}
Our understanding of extremely energetic processes in astrophysical systems critically relies on a combination of remote observations and numerical simulations. An excellent example of this synergy is represented by the problem of cosmic-ray acceleration. Observations tell us that Active Galactic Nuclei (AGN,~\cite{harrison2018agn}) can be a source of highly energetic particles. Emissions from blazar jets (AGN jets directed towards the observer) exhibit a high-energy component extending across the X-ray and gamma-ray bands together with a broad lower energy component~\cite{ulrich1997variability, fossati1998unifying, costamante2001extreme}. At the same time, the IceCube detector has traced high-energy neutrinos, produced by high-energy cosmic-ray protons, to a particular AGN, NCG 1068~\cite{IceCube2022_Sci}, which is compatible with the model of neutrino production in AGN coronae proposed by~\cite{inoue2020origin, kheirandish2021high, eichmann2022solving}. In both cases, magnetic reconnection~\cite{biskamp1996magnetic} is considered a viable process for the (pre-)acceleration of these high-energy particles. 
Numerical simulations can then help us verify this hypothesis.  

Magnetic reconnection converts the free energy contained in opposite-directed magnetic fields separated by a current sheet into bulk flows, plasma heating, and non-thermal high-energy particles, and occurs in a number of heliospheric and astrophysical environments,  see~\cite{zelenyi1990, zweibel2009magnetic, yamada2010magnetic, pucci2024applications} and the recently published collection~\cite{burch2025magnetic}. This process begins when either collisional or kinetic effects break the magnetic topology, producing small magnetic islands via the tearing instability \cite{Furth1963, Zelenyi1979, ji2011phase, pucci2013reconnection, Schoeffler2025}. The energy continues to be converted as the magnetic field is reconnected, while magnetic islands grow and merge. Energetic particles are accelerated via reconnecting electric fields as well as the Fermi process as they bounce between contracting magnetic islands. While suprathermal particle acceleration is also observed in non-relativistic regimes, relativistic plasmas are particularly suited for the production of high-energy non-thermal particles~\cite{Drake2006, Sironi2014, guo2015particle,  Guo2016, Werner2017, Werner2017b, petropoulou2018steady}.

One key point in studying the viability of magnetic reconnection as a (pre)-acceleration process in Active Galactic Nuclei (AGN) jets and coronae is the capability of a) simulating magnetic reconnection in magnetic field configurations resembling those found in AGNs and b) at length-scales compatible with AGNs. These two points are in fact related: simulating `realistic' magnetic field configurations that one expects in AGN jets and coronae requires addressing the huge separation between system scales, reconnection scales, and kinetic scales (if one intends to model particle acceleration self-consistently)~\cite{petropoulou2016blazar}. This second proposition is so challenging that most investigations of the topic of particle acceleration in AGN jets completely neglect the multi-scale nature of the problem, focusing instead either on its large-scale aspects via MHD simulations~\cite{bromberg2019kink, medina2021particle}, or on its kinetic aspects via fully kinetic Particle-In-Cell, PIC, simulations~\cite{Sironi2014, guo2015particle, Guo2016, Sironi2016, Werner2017, Werner2017b, alves2018efficient, petropoulou2018steady, petropoulou2019relativistic, Hoshino2020, Schoeffler2025}.
In this work, we present an attempt to increase the temporal and spatial scope of fully kinetic PIC simulations. As a first step towards bridging these scales: we explore whether a semi-implicit PIC discretization is an efficient option for fully kinetic, relativistic PIC simulations that model the onset of the tearing instability and the formation of suprathermal particle spectra. Concurrent activities in the same direction include benchmarking the spectra produced by test particles tracked through electromagnetic fields from two-fluid relativistic simulations of magnetic reconnection against what is obtained in fully kinetic PIC simulations~\cite{Wilbert2025}. Non-relativistic semi-implicit PIC discretizations, such as the implicit moment method~\cite{vu1992celest1d, lapenta2006kinetic}, have been validated long ago against the explicit PIC approach, see e.g.~\cite{ricci2004collisionless}, and are routinely used in non-relativistic simulations of magnetic reconnection (see, e.g.~\cite{innocenti2015evidence, lapenta2015secondary}) as a way of extending their purview in space and time. 

In relativistic pair plasmas,  acceleration by magnetic reconnection efficiently produces non-thermal power-law distributions of energetic particles with power laws that depend on the magnetization of the plasma in both 2D and 3D~\cite{Sironi2014, Sironi2016}. An extensive study on these power laws and their high-energy cutoffs investigated the dependence on the system size and the magnetization of the plasma in 2D in~\cite{Werner2016}.
Similar studies extended these results to electron--ion plasmas with varied mass ratios up to realistic values \cite{Guo2016, Werner2017}, and to plasmas composed of electrons, protons, and positrons~\cite{petropoulou2019relativistic}. These studies were also extended to 3D for pair plasmas \cite{Werner2017b} and for electron--ion plasmas \cite{Bacchini2025,werneruzdensky2024}.
In all cases, for sufficiently large system sizes, harder spectra were observed for increased magnetization, and the high-energy cutoff was proportional to the magnetization.

Reconnection is often relativistic in extreme astrophysical environments. Whether a plasma is in the relativistic regime is determined by the magnetization,
\begin{equation}
	\sigma_{c\alpha} = \frac{B_0^2}{4\pi n_\alpha  m_\alpha c^2},
\end{equation}
where $B_0$, $m_\alpha$, and  $n_\alpha$ are the upstream magnetic field, mass, and number density for the particle species $\alpha$. For ions, $\sigma_{ci}$ is therefore the ratio $c_A^2/c^2$ where $c_A$ is the Alfv\'en speed and $c$ is the speed of light. For high magnetizations $\sigma_{ci} > 1$, the Alfv\'en speed $c_A$ corresponds to the proper velocity at which an Alfv\'en wave would propagate. We have marked the magnetization with the subscript $c$ for ``cold'' to distinguish it from a so-called ``hot'' magnetization, which takes into account relativistic temperatures,
\begin{equation}
	\sigma_{h\alpha} = \frac{B_0^2}{4\pi n_\alpha h_\alpha},
\end{equation}
where $h_\alpha \approx 4 T_\alpha$ is the enthalpy for ultra-relativistic temperatures $T_\alpha/(m_\alpha c^2) \gg 1$. For non-relativistic temperatures the enthalpy $h_\alpha \approx m_\alpha c^2$, and therefore $\sigma_{c\alpha} = \sigma_{h\alpha}$.

In this work, we simulate both the linear tearing instability and nonlinear particle acceleration in the context of relativistic magnetic reconnection using semi-implicit Particle-in-Cell (PIC) code RelSIM~\cite{Bacchini2023}. We show that RelSIM reproduces results found in traditional explicit PIC simulations at significantly lower computational cost, and that these simulations can better model acceleration mechanisms responsible for several astrophysical observations.
Here, we lay out the organization of the paper. After this introduction in Section~\ref{sec:intro}, we will describe our methods, the setup of the simulations, the Harris equilibrium, and important length scales of the problem in Section~\ref{sec:setup}. In Section~\ref{sec:simparam} we describe the simulation parameters. We then explain our simulation results in Section~\ref{sec:results}, which is divided into two subsections; one for a set of runs verifying expected tearing instability behavior in the linear regime and one that tests the nonlinear stage of reconnection, where we measure non-thermal particle distributions. Finally, we will conclude with a discussion in Section~\ref{sec:conclusion} that highlights the connections of this study to astrophysical observations.
 
\section{Methods and Simulation Setup}\label{sec:setup}
In this work, we conduct Particle-in-Cell (PIC) simulations and take advantage of the Energy-Conserving Semi-Implicit Method ECSIM \cite{Lapenta2017, lapenta2017multiple,gonzalezherrero2018,gonzalezherrero2019,lapenta2023,croonen2024}, which has been extended to include relativistic effects in the Relativistic Semi-implicit Method RelSIM~\cite{Bacchini2023}. RelSIM can handle relativistic simulations while retaining very accurate energy conservation. The method is implicit only in the field solver while the particle pusher is explicit. The field solver uses a mass-matrix formulation that yields exact (to machine precision) energy conservation in the non-relativistic limit and considerably reduced energy errors in the relativistic case compared to a fully explicit scheme.
This allows for reduced resolution, which is not possible when using explicit models due to significant numerical heating and numerical instabilities. 

As a reference, we compare our results with simulations performed with the OSIRIS code \cite{OSIRIS}, which uses an explicit field solver, a Boris pusher, and a charge-conserving current deposition scheme \cite{Villasenor1992}. In our previous study~\cite{Schoeffler2025}, we have used OSIRIS simulations to verify the growth rate predictions in~\cite{Zelenyi1979} (including the ~\cite{Hoshino2020} correction) for the tearing instability in relativistic pair plasmas. In this paper, we extend our previous work to electron--proton plasmas (also covered in~\cite{Zelenyi1979}), where the scale separation between electrons and protons may make the usage of a semi-implicit scheme computationally advantageous.

All simulations are initialized in a double Harris kinetic equilibrium using the relativistic generalization ~\cite{KirkHarris} for relativistic temperatures ($T > m_e c^2/2$) with periodic boundary conditions.  The simulations are conducted in a physical domain ranging from $x = -L_x$ to $L_x$, and $y = -L_y$ to $L_y$, where $L_y$ is the distance between the two current sheets.  

Each current sheet consists of counter-drifting Maxwell--J\"uttner distributions of ions and electrons with a uniform temperature~$T$ in the rest frame of the respective species, boosted into opposite $\pm \hat{z}$-directions with a uniform speed~$v_d$. The drift speed $v_d$ corresponds to a Lorentz factor
$\Gamma_d = 1/\sqrt{1-v_d^2/c^2}$, and a proper drift speed $u_d = \Gamma_d v_d$. 
The lab-frame density profile (of both electrons and ions) in the Harris current sheet at $y = \pm L_y/2$ is given by
\begin{equation}
	n = n_0 {\rm sech}^2 \left(\frac{y \mp L_y/2}{a}\right) + n_b,
\end{equation}
where $n_0$ is the number density at the center of each current sheet for both ions and electrons. Since the diamagnetic drift determines the drift $v_d$, the magnitude of $u_{d}/T$ must be the same for both species.  An additional uniform background population $n_b$, that is at rest with temperature $T$, is included which does not disturb the kinetic equilibrium.

The pressure of the current sheets is balanced by self-consistent magnetic field profiles, which results in a kinetic equilibrium. The magnetic field profile is given by
\begin{equation}
	B_x =B_0\left[
	1 -\tanh\left(\frac{y - L_y/2}{a}\right)
	+\tanh\left(\frac{y + L_y/2}{a}\right)\right].
\end{equation}

The magnetic field can be calculated, using pressure equilibrium, to be
\begin{equation}
\label{forcebalance}
    B_0 = \sqrt{\frac{8 \pi (2 n_0) T}{\Gamma_d}},
\end{equation}
and using Amp\`ere's law, the current sheet half-thickness can be calculated to be
\begin{equation}
\label{ampere}
    a = \frac{c B_0}{4 \pi e (2 n_0) v_d} = \sqrt{\frac{T c^2}{2 \pi (2 n_0) e^2 \Gamma_d v_d^2 }} \approx \sqrt{\frac{\Gamma_{T\alpha}m_\alpha c^4}{4 \pi (2 n_0) e^2 \Gamma_d v_d^2 }}.
\end{equation}

In the relativistic regime, we express this in terms of the peak Lorentz factor $\Gamma_{T\alpha} \equiv 2T/(m_\alpha c^2)$ of a strongly relativistic Maxwell-J\"uttner distribution. Likewise, in the classical regime, we will use the thermal velocity $v_{T\alpha}$ ($v_{T\alpha}/c \equiv \sqrt{2T/(m_\alpha c^2)} $).

We can express the scales of the system, for species $\alpha$ ($i$ for ions or $e$ for electrons), as
the classical inertial length,
\begin{equation}
\label{classde}
    d_{0\alpha,C} = \sqrt{\frac{m_\alpha c^2}{4\pi n_0 e^2}},
\end{equation}
the relativistic inertial length,
\begin{equation}
\label{relde}
    d_{0\alpha,R} = \sqrt{\Gamma_{T\alpha}} d_{0\alpha,C} ,
\end{equation}
the classical Larmor radius,
\begin{equation}
\label{classrho}
    \rho_{L\alpha,C} = \frac{v_{T\alpha}}{\Omega_{c\alpha}} = \sqrt{\frac{\Gamma_d}{2}} d_{0\alpha,C} = \frac{u_d}{v_{T\alpha}}a,
\end{equation}
and the relativistic Larmor radius,
\begin{equation}
\label{relrho}
    \rho_{L\alpha,R} = \frac{\Gamma_{T\alpha} c}{\Omega_{c\alpha}} = \sqrt{\frac{\Gamma_d}{2}} d_{0\alpha,R} = \frac{u_d}{c}a,
\end{equation}
where $\Omega_{c\alpha} = e B_0/(m_\alpha c)$ is the cyclotron frequency. We also define the nominal Larmor radius $\rho_{0\alpha} \equiv c/\Omega_{c\alpha}$. Our constraint from force balance, \eq{forcebalance}, implies $\rho_{L\alpha} \approx d_{0\alpha}/\sqrt{2}$ in both classical and relativistic regimes
as seen in \eqs{classrho}{relrho} as long as $\Gamma_d\sim1$ (or upon redefinition using a temperature $T^\prime=T/\Gamma_d$). We do not precisely define $\rho_{L\alpha}$ in the transition between the classical and relativistic regimes, at $T/(m_\alpha c^2) \sim 1$ when $\rho_{L\alpha,C} \sim \rho_{L\alpha,R}$. We will therefore specify in the text when we use $\rho_{L\alpha,C}$ or $\rho_{L\alpha,R}$.

Until now, inertial lengths have been written in terms of $n_0$. Our standard definition of the inertial length will be in terms of the background density $n_b$, and thus
\begin{equation}
    \label{classdeb}
    d_{\alpha,C/R} = d_{0\alpha,C/R} \sqrt{\frac{n_0}{n_b}} = \rho_{L\alpha,C/R} \sqrt{\frac{\sigma_{c\alpha}m_\alpha c^2}{2T}}.
\end{equation}

Likewise, the plasma frequency $\omega_{pe} = c/d_{e,C}$ ($\omega_{pe0}= c/d_{0e,C}$) is defined in terms of $n_b$ ($n_0$). 

Using this setup, we will compare our results to the theoretical growth rate of the tearing instability from \cite{Zelenyi1979} taking into account modifications from \cite{Hoshino2020}. In the regime where the ion and electron populations have the same temperature $T$, but the thermal velocities are considered non-relativistic for the ions ($T/m_ic^2 \ll 1$) and relativistic for the electrons ($T/m_ec^2 \gg 1$), the theoretical growth rate is the following.
\begin{equation}
\label{sec:tearing}
\frac{\gamma_{th} a}{c} = \frac{2\sqrt{2}}{\pi} f(k) \left(\frac{u_d}{c}\right)^{3/2}\frac{1}{\Gamma_d^{5/2}}  
\frac{2}{1 + 
2^{7/4}\pi^{-1/2}(T/m_ic^2)^{1/4}}
\end{equation}
where
\begin{equation}
\label{sec:tearingdisp}
f(k) = ka(1-k^2a^2)
\end{equation}
with a fastest growing mode at $ka = 1/\sqrt{3}$.

\section{Simulation Parameters}\label{sec:simparam}

We will first test the linear behavior of the tearing instability for relativistic temperatures by comparing RelSIM results against linear theory and OSIRIS simulations. We will then examine the instability's nonlinear evolution and the subsequent development of non-thermal energy spectra. Here, we will run simulations with RelSIM only, and compare against results in~\cite{Guo2016} and~\cite{Werner2016}.

\subsection{Relativistic tearing}
For the simulations of the linear tearing instability, we choose a mass ratio $m_i/m_e = 100$, a temperature $T/(m_e c^2) = 10$ for both species, a half thickness of the current sheet $a/\rho_{Le,R} = 20$ ($a/\rho_{Li,C} = 8.94$), and a proper drift velocity $u_d/c = 0.05$. We do not consider background plasma ($n_b=0$) for this case. We can thus compare our results with the predictions from \cite{Zelenyi1979} where $a/\rho_{Li,C}> 1$ using the same methods as described in \cite{Schoeffler2025}.

For our OSIRIS simulations, we choose a length $L_x = 250 \rho_{Le,R}$, and $L_y = 255 \rho_{Le,R}$, with a resolution $dx = 0.625 \rho_{Le,R} = 0.0312 a$, and a time step $dt \omega_{pe0}= 0.758$ to resolve the electron plasma frequency. This corresponds to $dt c/dx = 0.384 < 1/\sqrt{2}$, which also satisfies the Courant condition.
We run each simulation for a few $e$-folding times ($t\gamma_{th}\sim 3$), calculated in terms of the theoretical growth rate from \eq{sec:tearing}. We compare 2 sets of simulations, varying the number of particles per cell between 256 and 4096.

We compare the OSIRIS simulations with a set of RelSIM simulations using the same physical parameters and box size as in OSIRIS.
For our numerical parameters, we consider a fiducial case with 256 particles per cell. The spatial resolution for this fiducial case is $dx = 5 \rho_{Le,R} = 0.25 a$, and the temporal resolution is the same as OSIRIS, $dt \omega_{pe}= 0.758$. In a first set of simulations, we hold the time resolution constant.
We perform 3 simulations, varying the number of particles per cell between $256, 1024,$ and $4096$.
We vary the resolution in space by ($dx/\rho_{Le,R} = 2.5, 5, 10$, and $20$; $dx/a = 0.125, 0.25, 0.5$ and $1$), keeping the time resolution and particles per cell constant.
In a second set of simulations, we hold the space resolution constant at $dx = 5 \rho_{Le,R}$, and then vary the resolution in time ($dt\omega_{pe0} = 0.379, 0.758, 1.516, 3.032,$ and $6.064$.). All these simulations are listed in Table~\ref{table}.

\subsection{Nonlinear reconnection}

In the RelSIM nonlinear simulations, we consider a background density $n_b \ne 0$ determined by $\sigma_{ce}$, to provide a constant supply of upstream plasma in the nonlinear stage, and a smaller half thickness of the current sheet $a/d_{e,C} = 0.5$ in order to reach the nonlinear stage more quickly.
We set the mass ratio to $m_i/m_e = 100$, the temperature to $T/(m_e c^2) = 10$ for both species, and the proper drift velocity to $u_d/c = 0.4$.

We choose a resolution of $dx = 1.6 d_{e,C} = 0.37d_{e,R} = 0.16 d_{i,C}$ so that we resolve both the ion and electron background inertial scales.  We use a time step $dt \omega_{pe} = 0.96$, which satisfies the Courant condition ($dt c/dx = 0.66 < 1/\sqrt{2}$). Note that in this regime, a typical thermal particle moves close to the speed of light. We use 16 and 256 particles per cell in the respective background and Harris sheet populations. Although we do not resolve the initial current sheet half thickness $dx = 3.34 a$, this current sheet is transient and is quickly replaced by new current sheets from the background population. We run each simulation for several light crossing times, up to $tc/L_y = 3.81$ or $7.77$.

For our fiducial case, we set the magnetization to $\sigma_{ci} = 100$ implying $n_0/n_b = 269.3$ ($\sigma_{he} = 250$), and use a length $L_x = L_y = 50 d_{i,C} ~(500 d_{e,C})$.
As previously stated, the initial high-density current sheet (a result of the high magnetization), which balances the magnetic pressure, is transient and is quickly replaced by lower-density background plasma.
In this fiducial case, $dx = 166 \rho_{0e} = 8.35 \rho_{Le,R} = 1.6 \rho_{0i} = 3.73 \rho_{Li,C}$, so we do not resolve the background electron or ion Larmor radius. However, if we assume that the typical temperature reaches the point where the ions move at the background Alfv\'en speed, the Larmor radius is well resolved for both species,
$dx =  0.16 \sqrt{\sigma_{ci}} \rho_{0,i} = 0.16 \sqrt{\sigma_{ci}} (m_i/m_e) \rho_{0,e}$.
When we vary $\sigma_{ci}$ the resolution of this Larmor radius $\sqrt{\sigma_{ci}} \rho_{0,i}$ remains fixed. 

To compare with the results in \cite{Guo2016} and \cite{Werner2016} who checked the dependence of the energetic particle spectra on system size and magnetization, we modify our fiducial case by varying these parameters. We vary the system size as $L_x/d_{i,C} = L_y/d_{i,C} = 50,100,$ and $200$ holding $\sigma_{ci} = 100$ constant (parameters from \cite{Guo2016}), which corresponds to the normalized system size from \cite{Werner2016} of $2L_y/\sigma_{ce}\rho_{0e}=10, 20,$ and $40$. We also vary the magnetization as $\sigma_{ci} = 10$, $100,$ and $1000$, holding $L_y/d_{i,C} = 50$, the half thickness of the current sheet $a/d_{e,C} = 0.5$, and the drift velocity $u_d/c = 0.4$ constant, which correspond to $2L_y/(\sigma_{ce}\rho_{0e})=31.6, 10,$ and $3.16$ respectively. These system sizes and magnetizations are also parameters taken from \cite{Guo2016}.

\section{Simulation results}\label{sec:results}
In this section, we examine two scenarios. First, we look at the linear stage of reconnection, considering no background population, and measure the tearing growth rates. We compare measurements from semi-implicit particle-in-cell simulations using RelSIM~\cite{Bacchini2023} with theoretical growth rates from \cite{Zelenyi1979} and measurements from explicit simulations using the OSIRIS framework~\cite{OSIRIS}.
Second, we look at the late-time nonlinear evolution with a background population included and measure the non-thermal particle spectra in simulations using RelSIM to compare with the results in \cite{Guo2016} and \cite{Werner2016}.

\subsection{Relativistic tearing}
Here we present and compare simulation results using both OSIRIS and RelSIM with analytical predictions from \cite{Zelenyi1979} using the methods described in \cite{Schoeffler2025}. We perform simulations using the simulation parameters described in Section~\ref{sec:setup}. The aim here is to show that RelSIM performs comparably well to explicit simulations while retaining a computational advantage.
\begin{table}
  \begin{center}
\def~{\hphantom{0}}
  \begin{tabular}{lccccccccc}
    sim\# & code & ppc & $dx/\rho_{Le,R}$ & $dt c/\rho_{Le,R}$ & $\gamma_m a/c$   & $k_{\rm{max}} a$ & $t_{\rm{nl}}\gamma_{th}$ & $|E-E_0|/E_0$\\[3pt]
       1 & \rm{OSIRIS} & ~256 & 0.625 & 0.756 & 0.0039  & 0.3 & 3.45 & $2\times10^{-2}$\%\\
       2 & \rm{OSIRIS} & 4096 & 0.625 & 0.756 & 0.0029  & 0.4&--& $1\times10^{-3}$\%\\
       3&\rm{RelSIM}   & ~256 & 5 & 0.756 & 0.0046  & 0.4&1.85& $6\times10^{-5}$\%\\
       4&\rm{RelSIM}   & 1024 & 5 & 0.756 & 0.0035  & 0.1&3.35& $2\times10^{-5}$\%\\
       5&\rm{RelSIM}   & 4096 & 5 & 0.756 & 0.0032  & 0.4&2.85& $5\times10^{-6}$\%\\
       6&\rm{RelSIM}   & ~256 & 20 & 0.756 & 0.0004  & 0.1&1.90& $7\times10^{-5}$\%\\
       7&\rm{RelSIM}   & ~256 & 10 & 0.756 & 0.0028  & 0.1&1.91& $1\times10^{-4}$\%\\
       8&\rm{RelSIM}   & ~256 & 2.5 & 0.756 & 0.0045  & 0.1&1.83& $8\times10^{-5}$\%\\
       9&\rm{RelSIM}   & ~256 & 5 & 6.064 & 0.0034  & 0.1&2.19& $3\times10^{-2}$\%\\
       10&\rm{RelSIM}   & ~256 & 5 & 3.032 & 0.0037  & 0.1&1.97& $1\times10^{-3}$\%\\
       11&\rm{RelSIM}   & ~256 & 5 & 1.516 & 0.0037  & 0.1&2.02& $2\times10^{-4}$\%\\
       12&\rm{RelSIM}   & ~256 & 5 & 0.379 & 0.0053  & 0.1&1.60& $8\times10^{-7}$\%
  \end{tabular}
  \caption{Set of relativistic tearing simulations, for which $T/(m_ec^2) = 10$, $m_i/m_e = 100$, and $a/\rho_{Le,R}=20$; therefore the theoretical growth rate $\gamma_{th} a/c = 0.0037$ at $ka=k_{\rm{max}}a\approx 0.5$; see \eq{sec:tearing}.  Included in the table is the simulation code used, number of particles per grid cell, measured growth rate $\gamma_m a/c$ after a low pass filter ($ka < 1$), fastest growing wave number $k_{\rm{max}}a$, and the time when the simulation reaches the nonlinear stage $t_{\rm{nl}}$.  In addition, we include the maximum percentage error in energy conservation by $t\gamma_{th} = 1$.  The growth rates are measured from a best fit between $t\gamma_{th} = 0.5$ and $1.5$  for most runs except Simulation 2, which is measured between $t\gamma_{th} = 1.0$ and $2.0$ and Simulation 12, which is measured between $t\gamma_{th} = 0.5$ and $1.0$.}
  \label{table}
  \end{center}
\end{table}
\begin{figure}
    \includegraphics[width = 1\textwidth]{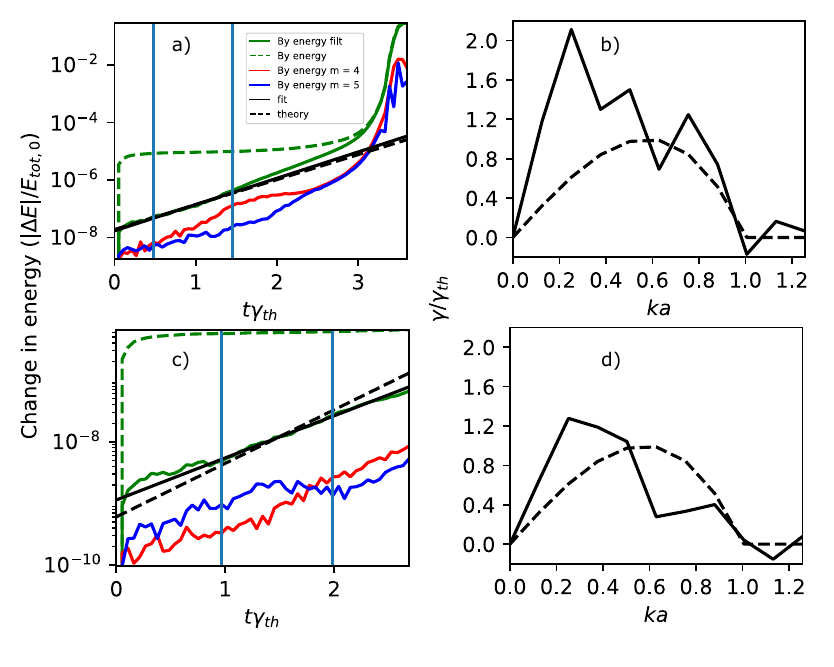}
    \caption{Evolution of the total energy in $B_y$ (dashed green), the low pass filtered energy (green), its fit (black), the theoretical growth rate from \eq{sec:tearing} (dashed black), and the energy in the $m=4$ (red) and $m=5$ (blue) modes, and the dispersion relation calculated as the best fit for the growth of each mode (black), and the theoretical dispersion from \cite{Zelenyi1979} and \eq{sec:tearingdisp} (black dashed). The top plots are the OSIRIS simulation with 256 particles per cell (Simulation 1 in Table~\ref{table})
    and bottom with 4096 particles per cell (Simulation 2).}
    \label{fig:tearinggrowthOS}
\end{figure}

We start with an OSIRIS simulation, with standard parameters; $dt \omega_{pe0} = 0.758$, $dx/\rho_{Le,R} = 0.625$, and $256$ particles per cell (Simulation 1 in Table~\ref{table}). When we examine the evolution of the energy, we see that the energy distribution remains relatively constant until reaching the fast-growing nonlinear stage shown in \cite{Schoeffler2025}, when the magnetic energy is rapidly converted into kinetic energy of the particles. The total energy in this simulation is conserved up to a maximum error of $2\times10^{-2}$\% by $t\gamma_{th} = 1.0$, a representative value before the fast-growing nonlinear stage.
The time here is normalized to the theoretical growth rate $\gamma_{th} a/c \approx 0.0037$ for non-relativistic ion temperatures and ultra-relativistic electron temperatures from \eq{sec:tearing}.
The noise in the $B_y$ component of the magnetic energy, the green dashed curve of {\bf Fig.~\ref{fig:tearinggrowthOS}a,} is too large to calculate a linear growth rate.
However, the green solid curve shows the energy after applying a low-pass filter of $B_y$ ($ka < 1$). We thus remove a significant portion of the noise, and can measure the growth rate between $0.5$ and $1.5$ $e$-folding times ($t\gamma_{th}=0.5-1.5$), finding $\gamma a/c \approx 0.0039$, consistent with theory. We use this range as a standard for measuring growth rates throughout this section.
We plot the evolution of the total magnetic field energy in $B_y$, the low-pass filtered energy, and the energy in two dominant individual modes in $k$-space ($m=4$ and $5$, where $k = \pi m/ L_x$ with $m\in\mathbb{N}$) in {\bf Fig.~\ref{fig:tearinggrowthOS}a.}.
The dominant mode at $1.5$ $e$-folding times is $m=4$, which corresponds to $ka \approx 0.5$, consistent with theory. The fastest growing mode, on the other hand, is at ($m=2$), growing at $ka \approx 0.3$ with a growth rate $\sim 2\gamma_{th}$ (see {\bf Fig. \ref{fig:tearinggrowthOS}b.}). As we have shown in \cite{Schoeffler2025}, as the growth enters the nonlinear stage (between the linear stage and the fast-growing nonlinear stage), the fastest growing mode saturates, and moves to lower wavenumbers. A theoretical evolution of the growth rate as the modes merge has been developed in \cite{Zeleny1988}.
One should note that to obtain a more reliable growth rate, more $e$-folding times are needed for the linear modes to have significant time to grow.
If the linear modes interact with each other, as is common as they reach the nonlinear stage, we expect errors in the measured growth rate.

We have therefore also performed a simulation with more (4096) particles per cell (Simulation 2 in Table~\ref{table}). We measure a growth rate between $1.0$ and $2.0$ $e$-folding times with reasonable agreement with theory: slightly less than the theoretical value ($\gamma a/c \approx 0.0029$; see {\bf Fig. \ref{fig:tearinggrowthOS}c.}).  
At this point, $ka \approx 0.4$ is the dominant mode, but it has begun to saturate. 
Looking at the growth rate of each mode as a function of wavelength, we see that again the fastest growing mode ($ka \approx 0.3$) is at a slightly lower wavenumber than the linear prediction, consistent with the beginning of the nonlinear stage (see {\bf Fig.~\ref{fig:tearinggrowthOS}d.}). Nevertheless, the errors in the measured growth rates of each mode have greatly improved. One could increase the particles per cell even further for a more reliable measurement, but this becomes computationally unwieldy.

\begin{figure}
    \includegraphics[width = 1\textwidth]{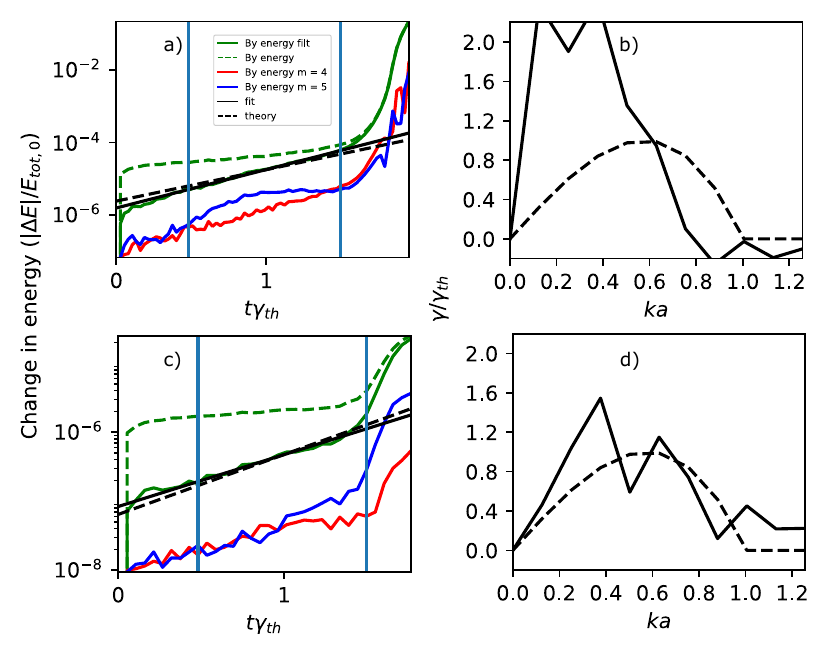}
    \caption{Evolution of the total energy in $B_y$ (dashed green), the low pass filtered energy (green), its fit (black), the theoretical growth rate from \eq{sec:tearing} (dashed black), and the energy in the $m=4$ (red) and $m=5$ (blue) modes, and the dispersion relation calculated as the best fit for the growth of each mode (black), and the theoretical dispersion from \cite{Zelenyi1979} and \eq{sec:tearingdisp} (black dashed). The top plots are from the RelSIM simulation with 256 particles per cell and 1/8 resolution in space compared to the OSIRIS simulations (Simulation 3 in Table~\ref{table}), and the bottom plots with 4096 particles per cell (Simulation 5).}
    \label{fig:tearinggrowthRS}
\end{figure}
We then look at RelSIM simulations with the same parameters, but taking advantage of the semi-implicit method. We consider a coarser resolution
in space by a factor of 8 along each direction ($dx/\rho_{Le,R} = 5$), keeping the temporal resolution constant. While the semi-implicit method tends to be slower by a factor of close
to $16$ in this configuration, we have a computational savings of a factor of $8^2$ due to the lower resolution. For our fiducial run (Simulation 3 in Table~\ref{table}), we find a growth rate that matches the
theory quite well, albeit slightly greater than the theoretical value ($\gamma a/c \approx 0.0046$; see {\bf Fig.~\ref{fig:tearinggrowthRS}a.}). Note that OSIRIS also overestimates the growth rate with low ppc, see {\bf Fig.~\ref{fig:tearinggrowthOS}a.}. The dominant wavelength is $ka \approx 0.3$, again a sign of already reaching the nonlinear regime
(see also {\bf Fig.~\ref{fig:tearinggrowthRS}b.}).
The particles per cell can be increased by a factor of $4$, from 256 to 1024, and we still observe a slight computational advantage.
We do this in Simulation 4 of Table~\ref{table} and find that the growth rate better matches the theory ($\gamma a/c \approx 0.0035$). Again, at around $1.5~e$-folding times, the fastest growing wave number $ka \approx 0.2$ is slightly lower than the linear prediction, consistent with nonlinear effects. The dominant mode, however, is at $ka \approx 0.4$, which is close to the linear prediction. 
For a more reliable measurement, one must increase the particles per cell even further. We therefore performed a simulation with $4096$ particles per cell (Simulation 5 in Table~\ref{table}); the growth rate still matches the theory ($\gamma a/c \approx 0.0032$; see {\bf Fig.~\ref{fig:tearinggrowthRS}c.}), now with the fastest modes growing close to $ka=0.5$ as expected by the theory (see {\bf Fig.~\ref{fig:tearinggrowthRS}d.}).
Like in {\bf Fig.~\ref{fig:tearinggrowthOS}., the lower noise in the simulation with higher particles per cell keeps the amplitude of the individual modes small, helping avoid interactions between the linear modes. Thus, we see good agreement in the measured growth rate for each mode.}

\begin{figure}
    \includegraphics[width = 1\textwidth]
    {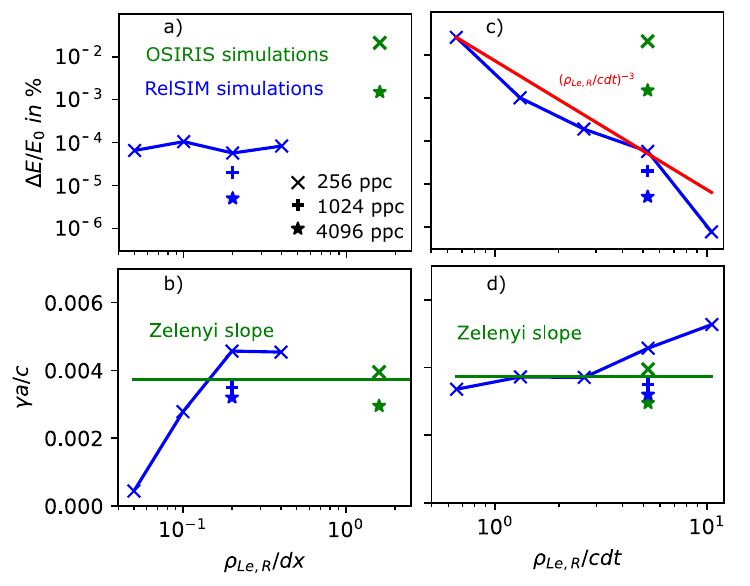}
    \caption{Scaling of the energy conservation (a,c) and the growth rate (b,d) as a function of the spatial resolution (a,b) and time resolution (b,c). The theoretical growth rate from \cite{Zelenyi1979},  i.e. \eq{sec:tearing}, is included as a green solid line. The measured growth rates from the OSIRIS Simulations (green) and the RelSIM Simulations (blue) with 256 ppc are indicated by crosses. Higher particle-per-cell simulations are indicated by plus signs for 1024 ppc (Simulation 4), and stars for 4096 ppc (Simulations 2 and 5). The growth rate has a weak dependence on the time resolution, and the energy conservation has a strong dependence $\sim(\rho_{Le,R}/cdt)^{-3}$.}
    \label{fig:scaling}
\end{figure}
To test the limits of the method, we perform a parameter scan for our RelSIM test case with $256$ particles per cell, varying both time and spatial resolution.
First, we vary the space resolution with $dx/\rho_{Le,R} = 2.5, 5, 10, 20$, where the lowest resolution corresponds to $dx/a = 1$ (see Simulations 6--8 in Table~\ref{table}). While energy conservation remains constant at about $1\times10^{-4}$ percent after $1$ $e$-folding time (see {\bf Fig.~\ref{fig:scaling}a.}),
before entering the fast-growing nonlinear regime, for the lowest-resolution case, the measured growth rate is much lower than theory $(\gamma a/c \approx 4.4\times10^{-4})$ (see {\bf Fig.~\ref{fig:scaling}b.}). The measured growth rate converges to a value close
to the theoretical one around $dx/\rho_{Le,R} = 5 (\gamma a/c \approx 0.0046)$, validating our choice. 

Next, we vary the time resolution with $dt/\omega_p = 0.379, 0.758, 1.516, 3.032, 6.064$ (see Simulations 9--12 in Table~\ref{table}).
When increasing the time resolution, the energy conservation is improved significantly (see {\bf Fig.~\ref{fig:scaling}c.}). For the lowest resolution ($dt/\omega_p = 6.064$), the energy is conserved up to $2\times10^{-2}$ percent,
comparable to the OSIRIS Simulation 1, which is significantly better resolved in time ($dt \omega_{pe0}= 0.758$). For the next lowest resolution, the energy conservation improves to $1.0\times10^{-3}$ percent, which is already better than the OSIRIS Simulation 2 with $4096$ particles per cell, with $1.4\times10^{-3}$\%. For the highest resolution run, we achieve an energy conservation of $7\times10^{-7}$\%. However, when changing the resolution, the growth rate increases slightly from $\gamma a/c \approx 0.0033$ to $0.0053$, i.e. faster growth for higher resolution (see {\bf Fig.~\ref{fig:scaling}d.}). For the highest resolution, the growth rate must be calculated between $t\gamma_{th}= 0.5$ and $1$, since the growth rate is faster than theory and thus the nonlinear stage is reached faster.
We note that, as we have shown earlier for the fiducial case, using more particles per cell mitigates this effect.

The semi-implicit method is thus useful to model the tearing instability, and as we discuss in the next section, can also be used for the long-term evolution of reconnection. Long-term simulations using the explicit method are limited by numerical heating, which breaks the energy conservation. While not perfectly energy-conserving, RelSIM has a strong advantage over explicit methods, which do not conserve energy as well. To compare with the computationally expensive OSIRIS simulation with $4096$ particles per cell, one can run with a reduced time resolution of $dt \omega_{pe0} = 3.032$ (in addition to the reduced spatial resolution), and have the same degree of energy conservation. Taking into account the intrinsic slowdown associated with semi-implicit methods, this leads to a net computational advantage of a factor of $256$.

\subsection{Nonlinear reconnection and spectra}
\label{sec:reconnectionSpectra}
We can take advantage of the strong energy conservation of RelSIM to explore the nonlinear evolution of the tearing instability and reconnection, and the subsequent development of non-thermal energy spectra, avoiding the numerical heating commonly found in explicit simulations. Here, rather than comparing with OSIRIS simulations, we compare with the results in \cite{Guo2016} and \cite{Werner2016}.

\begin{table}
  \begin{center}
\def~{\hphantom{0}}
  \begin{tabular}{lcccccc}
    $L_y/d_{i,c}$  & $\sigma_{ci}$   &   $L_y/\rho_{0i} \sigma_{ci}$   &
    $\alpha$ & $u_{\rm{max}}/c\sigma_{ce}$ & $t_{\rm{max}}c/L_y$ & 
    $|E-E_0|/E_0$\\[3pt]
       ~50   & ~100  & ~5~~~ & -1.35& 0.4&7.77& 0.4\%~\\
       100  & ~100  & 10~~~ & -1.35& 0.6&7.77& 0.3\%~\\
       200  & ~100  & 20~~~ & -1.35& 0.7&3.81& 0.2\%~\\
       ~50   & ~~10& 15.8~ & -1.50& 0.5&7.77& 0.06\%\\
       *50   & ~100& ~5~~~ & -1.35& 0.4&7.77& 0.4\%~\\
       ~50   & 1000& ~1.58 & -1.10& 0.1&7.77& 0.4\%~\\
  \end{tabular}
  \caption{Set of nonlinear reconnection simulations, for which $T/m_ec^2 = 10$, $m_i/m_e = 10$, and $a/\rho_{Le,R}=0.5$.
	  Included is the distance between the current sheets $L_y/d_{i,c}$, cold ion magnetization $\sigma_{ci}$,
	  the normalized system size $L_y/\rho_{0i} \sigma_{ci}$, the cutoff on the electron momentum $u_{\rm{max}}/c\sigma_{ce}$,
	  and the total time of the simulation $t_{\rm{max}}$. 
  	  In addition, we include the percentage error in energy conservation at $t_{\rm{max}}$. 
	  The system size varies in the first three simulations, while the magnetization varies in the second three. * Note that this is the same simulation as the first.
  }
  \label{table2}
  \end{center}
\end{table}
\begin{figure}
    \includegraphics[width = 1\textwidth]{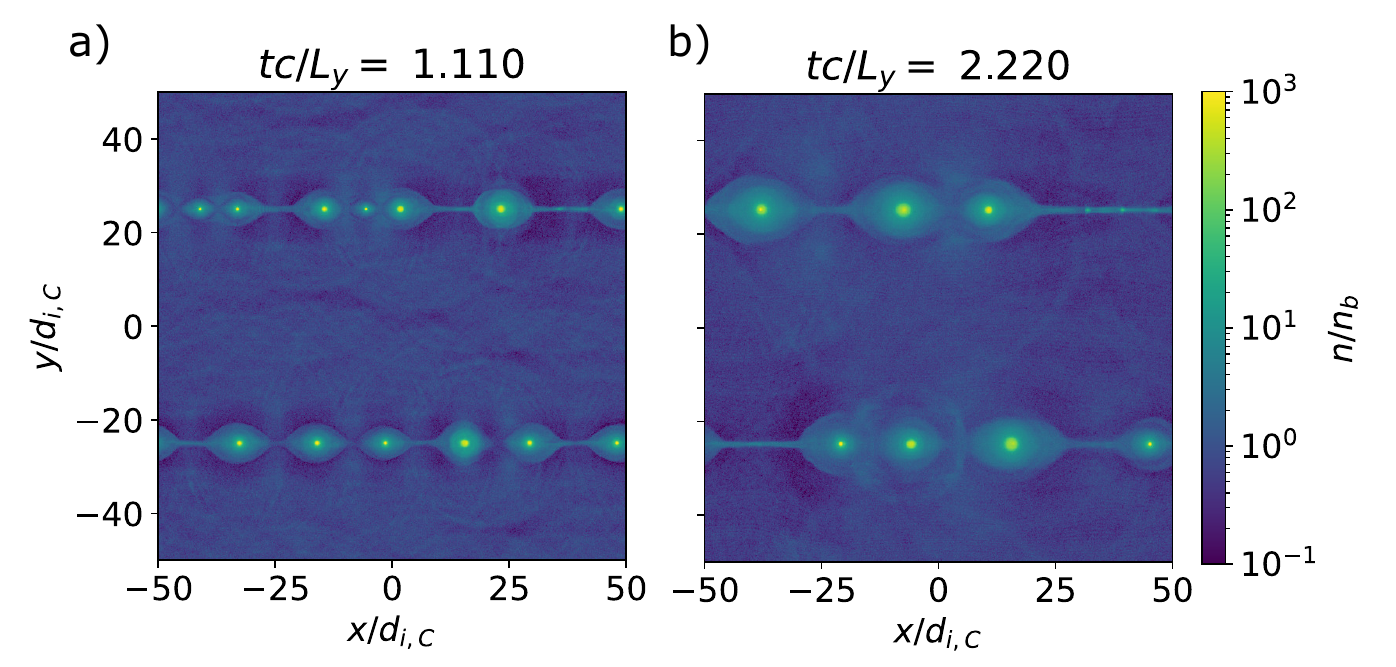}
    \caption{Map of normalized electron density $n/n_b$ at $t c/L_y = 1.11$ (a) and $2.22$ (b), showing the development of magnetic islands via tearing that eventually (over long times) merge until reaching the system size, for the fiducial simulation with $L_y/d_{i,C} = 50$ and $\sigma_{ci}=100$. }
    \label{fig:fiducialmap}
\end{figure}
\begin{figure}
    \includegraphics[width = 1\textwidth]{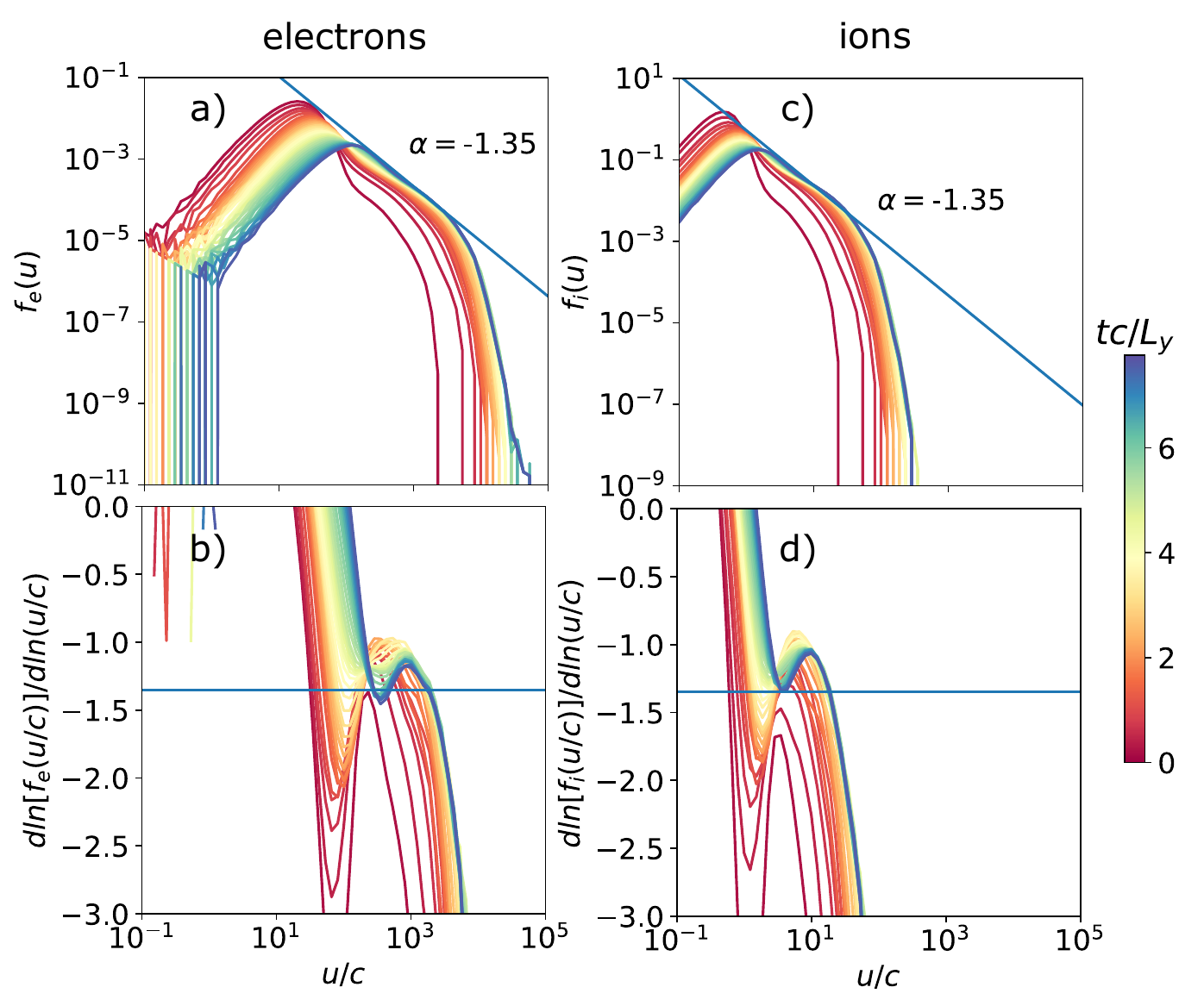}
    \caption{Spectra (top) and slope of the spectra (bottom) as a function of the proper speed $u/c$ of the electron (left) and ion (right) distributions of the background population for several times $t c/L_y$ for the fiducial case with $L_y/d_{i,C} = 50$ and $\sigma_{ci}=100$. The spectra fit a power-law slope $\alpha = -1.35$ between $u/c =100$ and $4000$ ($0.4 \sigma_{ce}$) for the electrons, and $u/c =1.5$ and $40$ ($0.4\sigma_{ci}$) for the ions. }
    \label{fig:ionspectra}
\end{figure}
First we will describe the results for our fiducial simulation. {\bf Fig.~\ref{fig:fiducialmap}.} shows the electron density of the reconnecting current sheets after about 2 light crossing times. Magnetic islands are generated via tearing and merge to the size of the box.
We examine the energy spectra in {\bf Fig.~\ref{fig:ionspectra}.}.
The peak of the spectra rises from the initial thermal Lorentz factor $u_{\rm{peak}}/c = \Gamma_{Te} = 20$ to $100$, and a power law is generated with spectral slope $\alpha=-1.35$, between $u/c = 100$ and $4000$ ($0.4\sigma_e$) for the electron distribution (see~{\bf Fig.~\ref{fig:ionspectra}a.}).
For the ions (see {\bf Fig.~\ref{fig:ionspectra}b.}, the peak of the spectra rises from the initial thermal velocity $u_{\rm{peak}}/c = v_T/c = 0.45$ to $1.4$, also with a power law $\alpha=-1.35$, between $u/c = 1.5$ and $40$ ($0.4\sigma_i$). The cutoff of the power-law slope is thus $u_{\rm{max}}/c \sim \sigma_\alpha$, which corresponds to the same energy for both electron and ion populations. This matches the observations in \cite{Guo2016}.

One should note that our simulations presented in {\bf Fig.~\ref{fig:ionspectra}. have been run to about $t \sim 7.5~L_y/c$, in comparison to \cite{Guo2016} who ran up to $t \sim 4~L_y/c$. Therefore, heating in the current sheets affected a large percentage of particles, explaining the increase of the peak Lorenz factor, particularly noticeable in our study. In realistic systems with open boundary conditions, this effect would be less pronounced as background particles are replenished and heated particles can escape.}

\begin{figure}
    \includegraphics[width = 1\textwidth]{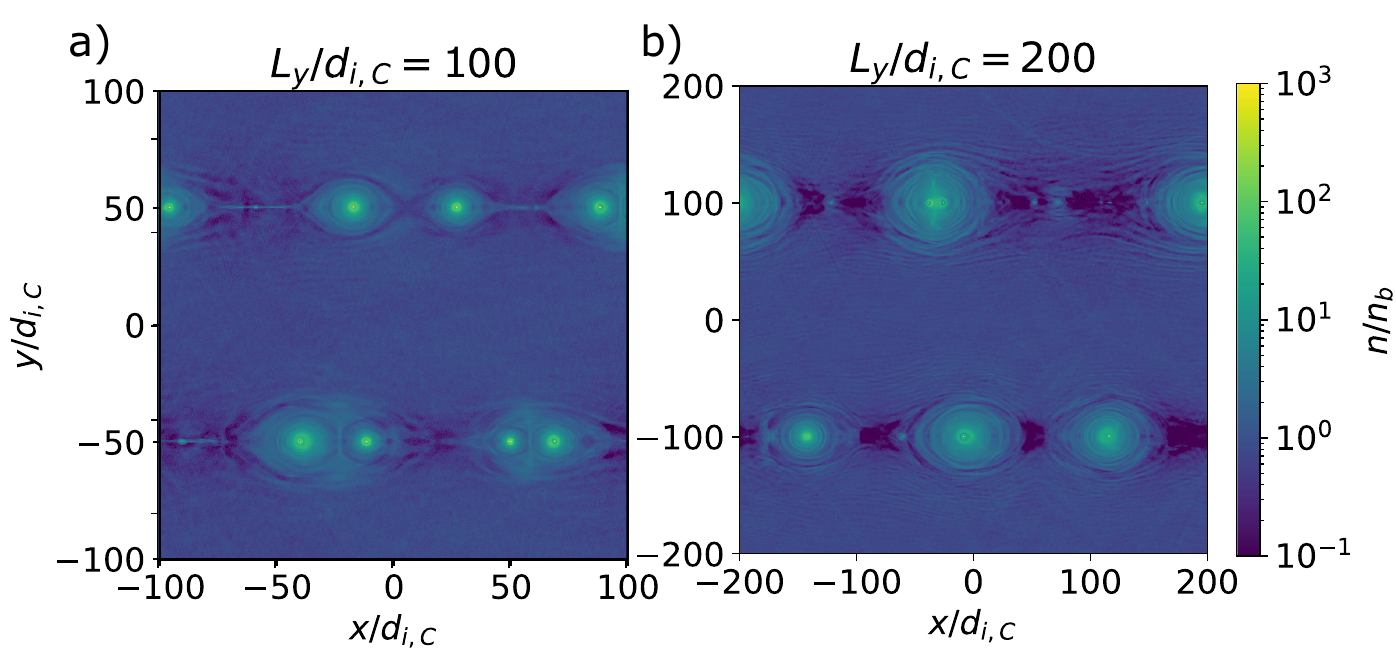}
    \caption{Maps of normalized electron density $n/n_b$ at $t c/L_y = 2.22$, showing the development of magnetic islands via tearing that eventually (over long times) merge until reaching the system size, for $L_y/d_{i,C} = 100$, and $200$ holding $\sigma_{ci}=100$ constant (like the fiducial case with $L_y/d_{i,C} = 50$). }
    \label{fig:Lmap}
\end{figure}
\begin{figure}
    \includegraphics[width = 1\textwidth]
    {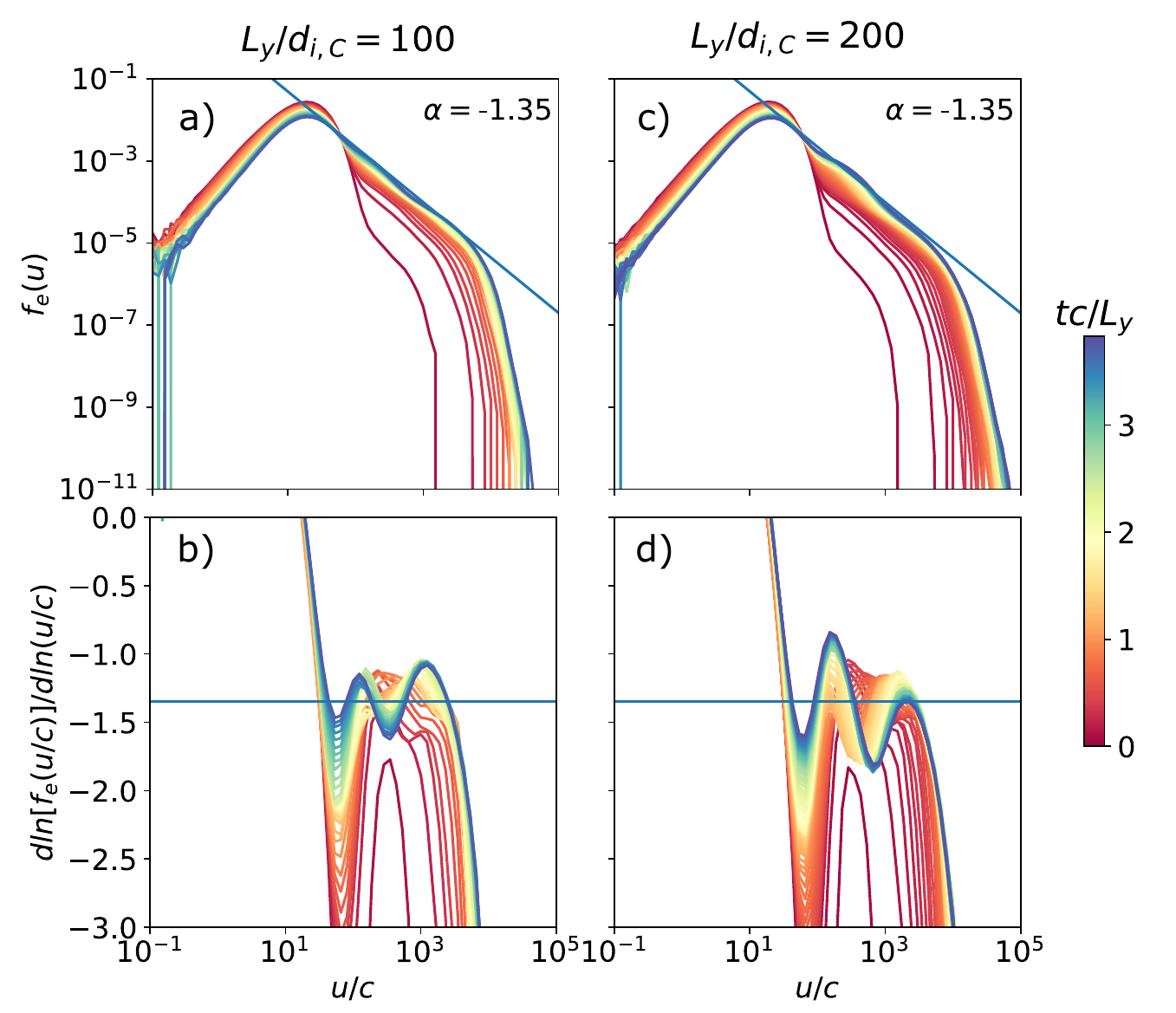}
    \caption{Spectra (top) and slope of the spectra (bottom) as a function of the proper speed/normalized momentum $u/c$ of the electron distribution of the background population for several times $t c/L_y$ for the cases $L_y/d_{i,C} = 100$ and $200$ holding $\sigma_{ci}=100$ constant (like the fiducial case with $L_y/d_{i,C} = 50$). The spectra fit a power-law slope $\alpha = -1.35$ between $u/c = 20$ and $6000$ ($0.6\sigma_{ce}$), and
    $u/c = 20$ and $7000$ ($0.7\sigma_{ce}$) for the respective cases.}
    \label{fig:Lspectra}
\end{figure}
We perform simulations for a total of 3 system sizes: $L_y/d_{i,C} = 50$, $100$, and $200$, including the fiducial case, which uses the same values used in \cite{Guo2016}. {\bf Fig.~\ref{fig:Lmap}.} shows that all cases are similar to the fiducial case; magnetic islands are generated via tearing and merge. Although not shown in the figure, they all eventually reach the size of the box. The thickness and length of the current sheets connecting the islands remain fixed to the kinetic scales.

If we look at the electron energy spectra for these systems in {\bf Fig.~\ref{fig:Lspectra}.}, we note that the power law persists with a constant $\alpha=-1.35$, independent of the system size, matching the observations in \cite{Guo2016}. Unlike the fiducial case, we only run until $t \sim 4~L_y/c$, such that the bulk heating is relatively insignificant. At any rate, in larger systems, more of the energy goes to the non-thermal power-law component. While the peak of the spectra remains close to $u_{\rm{peak}}/c = \Gamma_{Te}$, and the spectral slope is similar, the cutoff has a slight dependence on system size, occurring around $u_{\rm{max}}/c\sigma_{ce} = 0.4$, $0.6$, and $0.7$ for the increasing system sizes, and converging around $u_{\rm{max}}/c = \sigma_{ce} = 10^4$. This leads to a wider spectral range that fits a power law for larger system sizes.
\begin{figure}
    \includegraphics[width = 1\textwidth]{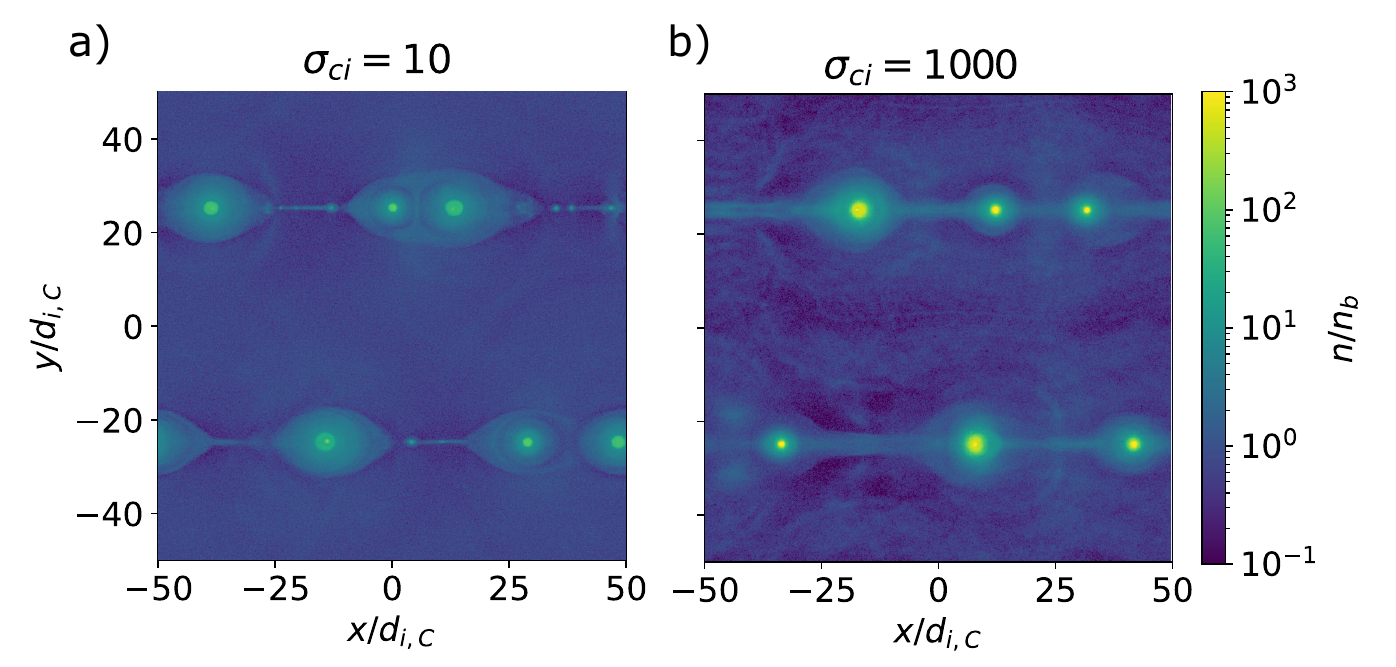}
    \caption{Maps of normalized electron density $n/n_b$ at $t c/L_y = 2.22$, showing the development of magnetic islands via tearing that eventually (over long times) merge until reaching the system size, for $\sigma_{ci} = 10$ and $1000$ holding $L_y/d_{i,C}=50$ constant (like the fiducial case with $\sigma_{ci} = 100$).}
    \label{fig:sigmamap}
\end{figure}
\begin{figure}
    \includegraphics[width = 1\textwidth]{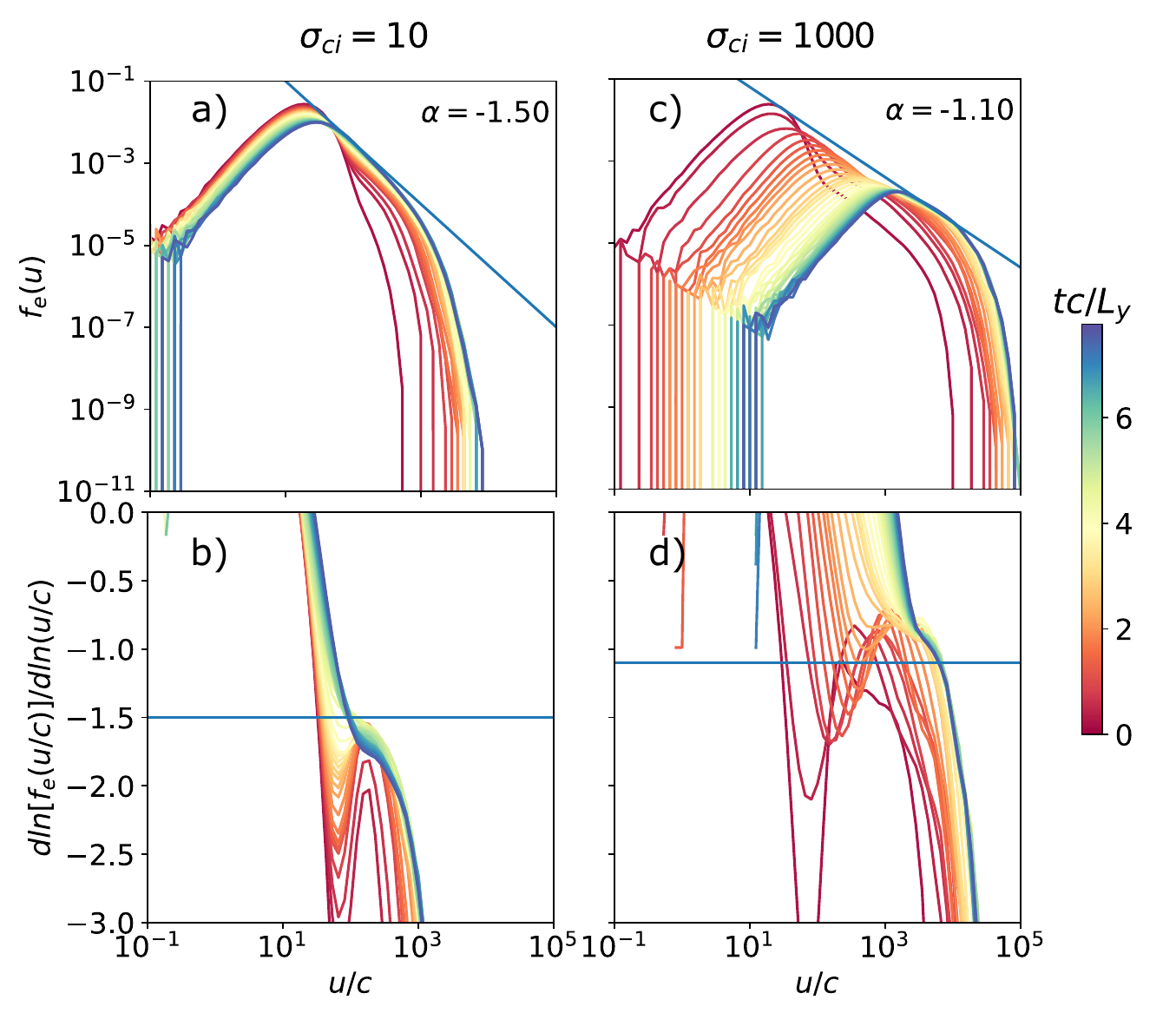}
    \caption{Spectra (top) and slope of the spectra (bottom) as a function of the proper speed $u/c$ of the background population electron distribution for several times $t c/L_y$ for the cases $\sigma_{ci} = 10$ and $1000$ holding $L_y/d_{i,C}=50$ constant (like the fiducial case with $\sigma_{ci} = 100$). The spectra fit a power-law slope $\alpha$ between $u/c = 30$ and $500$ ($0.5\sigma_{ce}$), and
    $u/c = 1000$ and $10000$ ($0.1\sigma_{ce}$) for the respective cases. }
    \label{fig:sigmaspectra}
\end{figure}

We also performed simulations for a total of 3 magnetizations, including the fiducial case: $\sigma_{ci} = 10$, $100$, and $1000$. {\bf Fig.~\ref{fig:sigmamap}.} shows the densities of the reconnecting current sheets for each simulation after about 2 light crossing times. As with the study of the system size, magnetic islands are generated via tearing and eventually merge, reaching the size of the box. The thickness and length of the current sheets depend on the magnetization. While we start with a current sheet with the same temperature as the background, the new current sheets that form have a temperature proportional to $\sigma_{ci}$, and a thickness that appears to scale with $\rho_{Li,R} \sim T$. The length of the current sheets is likely determined by a critical aspect ratio, and thus proportional to the thickness.

If we look at the electron energy spectra for these systems in {\bf Fig.~\ref{fig:sigmaspectra}.}, we note that the power law depends on the magnetization, with $\alpha=-1.5,-1.35, -1.10$ for $\sigma_{ci}=10$, $100$, and $1000$ respectively. This again matches the observations in \cite{Guo2016}. For more magnetized systems, there is a stronger bulk heating of the system. The peak $u_{\rm{peak}}/c$ of the distribution heats up to $30$, $100$, and $1000$, which converges to $u_{\rm{peak}}/c = \sigma_{ci}$.
The cases simulated here have a relatively small system size ($2L_y/(\sigma_{ce}\rho_{0e})=31.6, 10,$ and $3.16$). As we have shown for $\sigma_{ci} = 100$, a larger system size would likely reduce the start of the power law to $u_{\rm{peak}}/c = \Gamma_{Te}$.
The power-law cutoff occurs close to $u_{\rm{max}}/c = \sigma_{ce}$, but the factor in front decreases with magnetization: $u_{\rm{max}}/c \sigma_{ce} = 0.5,0.4,$ and $0.1$. This is likely because, for higher magnetizations, more energy goes to bulk heating. 

The widest spectral ranges, therefore, occur in sufficiently large systems with high magnetizations, where the required size is proportional to the magnetization ($2L_y/(\sigma_{ce}\rho_{0e}) \gg 1$). This is consistent with the dependence of the power-law cutoff on $2L_y/(\sigma_{ce}\rho_{0e})$ explained in \cite{Werner2016}. For high $2L_y/(\sigma_{ce}\rho_{0e})$ the cutoff depends on $\sigma_{ce}$,
while for low $2L_y/(\sigma_{ce}\rho_{0e})$ the cutoff depends on $L_y/\rho_{0e}$. Note that the same arguments are true for the ion spectral range; the transition length is the same $2L_y/(\sigma_{ci}\rho_{0i}) = 2L_y/(\sigma_{ce}\rho_{0e})$, and the cutoff occurs close to $\sigma_{ci}$.

\begin{figure}
    \includegraphics[width = 1\textwidth]{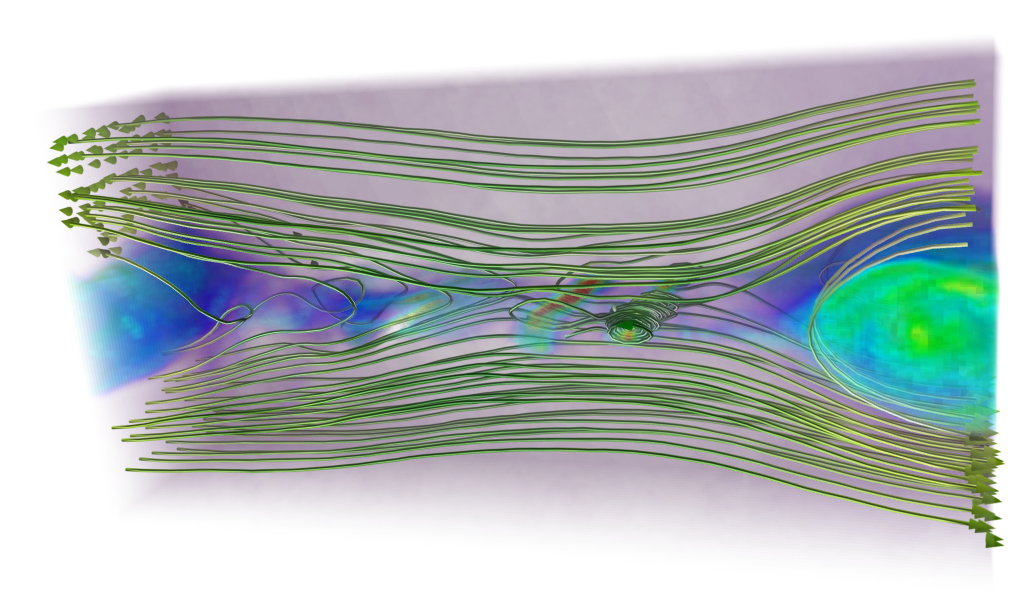}
    \caption{Rendering of electron density and magnetic field lines at $t c/L_y = 2.22$ for a 3D simulation with $L_y/d_{i,C} = 50$ and $\sigma_{ci}=100$.}
    \label{fig:density3D}
\end{figure}
\begin{figure}
    \includegraphics[width = 1\textwidth]{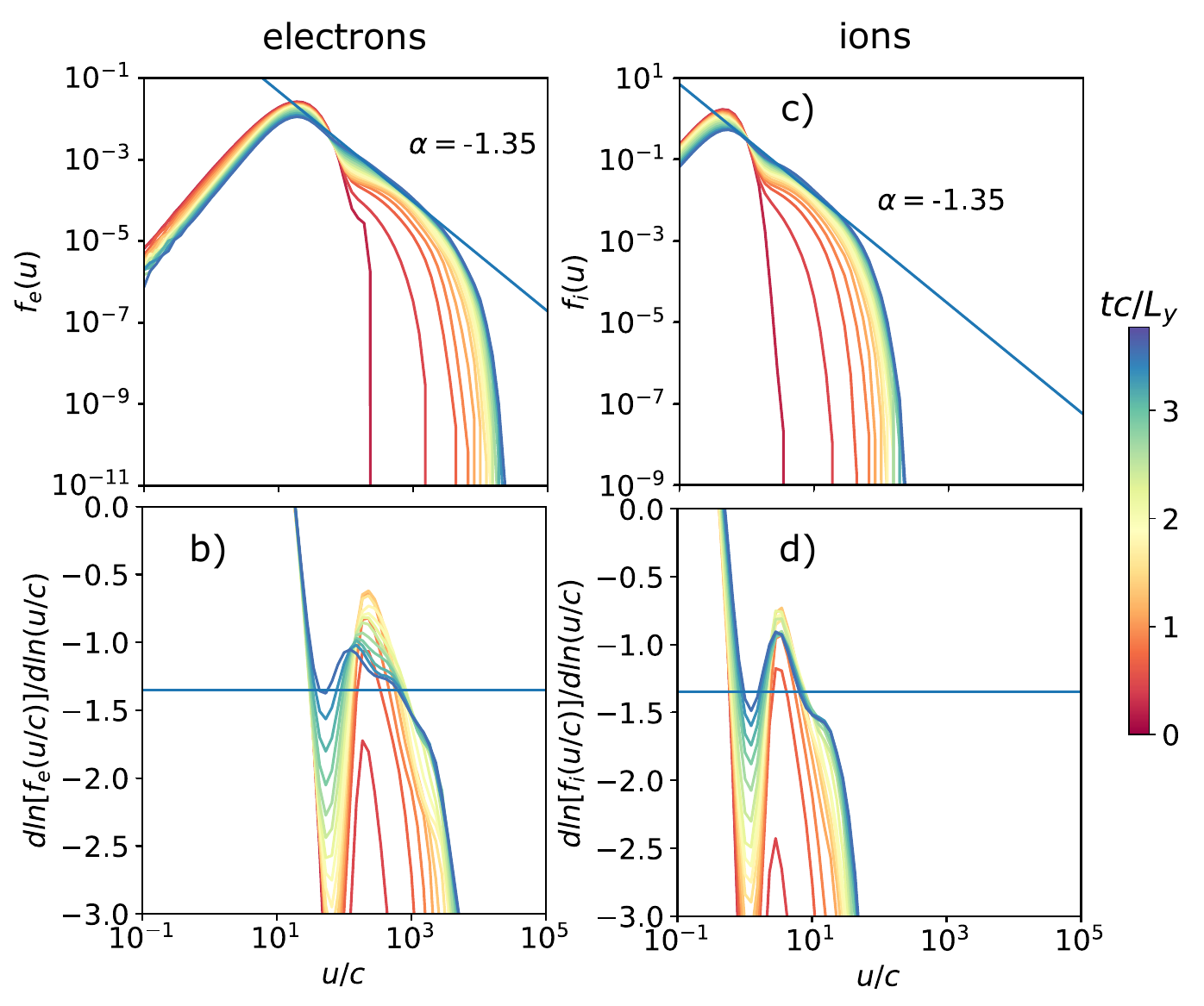}
    \caption{Spectra (top) and slope of the spectra (bottom) as a function of the proper speed $u/c$ of the electron (left) and ion (right) distributions of the background population for several times $t c/L_y$ for the 3D case with $L_y/d_{i,C} = 50$ and $\sigma_{ci}=100$. The spectra fit a power-law slope $\alpha = -1.35$ between $u/c =20$ and $2000$ ($0.4 \sigma_{ce}$) for the electrons, and $u/c =0.5$ and $40$ ($0.4\sigma_{ci}$) for the ions.}
    \label{fig:spectra3D}
\end{figure}
We finally perform one 3D simulation with parameters equivalent to the fiducial case, but extended in the z direction with $L_z = L_y/5$, using $8$ and $216$ ppc for the respective background and current sheet populations.
{\bf Fig.~\ref{fig:density3D}.} shows the rendered electron density with magnetic fields at $t c/L_y = 2.22$. This illustrates that the generation and merging of magnetic islands are reproduced in the 3D geometry. Furthermore, {\bf Fig.~\ref{fig:spectra3D}.} shows that the non-thermal distributions found in 2D are also reproduced in 3D.

We have thus run a set of simulations with relatively low resolution ($dx =  0.16 \sqrt{\sigma_{ci}} \rho_{0,i}$; $dx/\rho_{Le,R} = 2.6\mbox{--}26$) and with only $16$ particles per cell in the background, for as long as $7.77$ light crossing times with less than 1\% error in the energy (See Table~\ref{table2}).
The scaling of the spectral index as a function of the magnetization and system size index matches the results of both \cite{Guo2016} and \cite{Werner2016}, even though the initial setup from \cite{Guo2016} is a force-free current sheet instead of a Harris sheet as we use here and in \cite{Werner2016}. This demonstrates that it is therefore possible to use the computational advantages of the high energy-conservation semi-implicit RelSIM code to study the spectral indices of large magnetized systems.

\section{Conclusions and outlook}\label{sec:conclusion}
In this paper, we have demonstrated that using semi-implicit methods is a promising approach to model the tearing instability and magnetic reconnection in relativistic regimes of electron--proton plasmas, which are notoriously difficult to model due to the scale separation between macroscopic and kinetic scales. We have performed RelSIM \cite{Bacchini2023} simulations modeling the tearing instability starting from a Harris equilibrium, finding results that match both theory \cite{Zelenyi1979,Schoeffler2025} and explicit OSIRIS \cite{OSIRIS} simulations. In further RelSIM simulations, we have also shown subsequent reconnection and acceleration of non-thermal particles that lead to power-law spectra that match previous work with standard explicit PIC models, e.g.~\cite{Guo2016} and \cite{Werner2016}.
As semi-implicit simulations with RelSIM avoid instabilities and strong numerical heating present in explicit simulations, this is possible with reduced resolution in both space and time, while retaining a high conservation of total energy. We have shown that by taking into account the savings of these lower resolutions, we can run simulations with the same degree of energy conservation for a computational cost that is up to 256 times cheaper than an equivalent simulation using an explicit code like OSIRIS, and even more significant savings are expected in 3D.
We have thus presented one example simulation showing that the same results are possible using fully 3D kinetic simulations.
This computational benefit opens up the possibility to model and explore regimes that were previously not possible with explicit methods. Furthermore, recent efforts have enabled ECSIM to run efficiently on GPU systems \cite{GPUpresentation}. This implementation should soon facilitate the integration of our new methods with the latest and fastest high-performance computing systems.

Using semi-implicit methods thus provides the opportunity to model reconnection with a new level of detail, modeling the acceleration phenomena in a variety of astrophysical sources such as pulsar wind nebulae, gamma-ray bursts, or active galactic nuclei (AGN). Most recently, the groundbreaking high-energy neutrino signal observed by the IceCube experiment \cite{IceCube2022_Sci}, which is commonly believed to originate from the AGN corona of NGC\,1068 e.g. \cite{Eichmann2022}, indicated that cosmic-ray protons need to be accelerated up to $\Gamma_{Tp}\gtrsim 10^4$ in such sources. However, it is currently not understood which process actually drives the protons in the center of an AGN up to those energies. Since general relativistic magnetohydrodynamic simulations 
 \cite{Fromm2022, Olivares2023} show no evidence of the presence of shocks in the central region of an AGN, the most promising processes are currently relativistic reconnection e.g. \cite{Fiorillo2024} or/and stochastic acceleration \cite{Mbarek2024}. The application of the semi-implicit method can provide new insights into the relative contribution of these two acceleration mechanisms in the energization of cosmic-rays in a number of astrophysical systems. 

Taking advantage of the separation of scales between classical ions and relativistic electrons at smaller magnetizations $\sigma_{ce} \sim 1$, one can explore parameter spaces to find spectra and spectral cutoffs relevant for AGN modeling. See \cite{Ball2018,Meringolo2023,Fromm2022}.

Furthermore, we are now well-positioned to look at bigger system sizes, where models of reconnection, similar to what was done in this paper, include the effects of turbulence and stochastic acceleration \cite{Bacchini2025}.

The most promising regime to take advantage of the semi-implicit method is in large 3D systems, as one can benefit from the lower spatial resolution in each dimension. While we have shown that similar results to what we have shown in our 2D simulations can be obtained, a more careful study is still needed. Future studies would check the effects of the spatial extent in the $z$ direction, variation of physical parameters, and the influence of instabilities with wavenumber along the $z$ direction e.g. kinking instabilities.

\section*{Data availability}
The main data and input files supporting the findings of this study are openly available
in Zenodo at https://zenodo.org/records/16883720, reference number 16883720.

\section*{Code availability}
The version of OSIRIS used for simulations in this study is freely available as open source.
RelSIM is available on request after signing a user agreement.

\bibliographystyle{cip-v3-bst-submit}

\bibliography{implicitspectranew}

\section*{Funding}
This work is supported by the German Science Foundation DFG within the Collaborative Research Center SFB1491 and DFG project 544893192.
The authors gratefully acknowledge the Gauss Centre for Supercomputing e.V. (https://www.gauss-centre.eu/) for funding this project by providing computing time on the GCS Supercomputer SUPERMUC-NG at Leibniz Supercomputing Centre (https://www.lrz.de/).
Simulations were performed at SuperMUC (Germany).
F.B.\ acknowledges support from the FED-tWIN programme (profile Prf-2020-004, project ``ENERGY'') issued by BELSPO, and from the FWO Junior Research Project G020224N granted by the Research Foundation -- Flanders (FWO). 

\section*{Author contributions}
KMS performed all simulations and produced the main manuscript. FB provided semi-implicit code RelSIM and contributions to the manuscript. KK provided expertise on numerical methods and contributed to the manuscript. BE provided expertise on observations and astrophysical context and contributed to the manuscript. MEI made major contributions to the manuscript and references.

\section*{Competing interests}
There are no competing interests.

\end{document}